\documentclass[12pt]{article}

  \usepackage{graphicx}
  \usepackage{epsfig}
  \usepackage{amssymb}
\evensidemargin - .5truecm
\begin{document}

\newcommand{\be}{\begin{equation}}
\newcommand{\ee}{\end{equation}}
\newcommand{\nn}{\nonumber}
\newcommand{\bea}{\begin{eqnarray}}
\newcommand{\eea}{\end{eqnarray}}
\newcommand{\bfig}{\begin{figure}}
\newcommand{\efig}{\end{figure}}
\newcommand{\bc}{\begin{center}}
\newcommand{\ec}{\end{center}}
\def\ad{\dot{\alpha}}
\def\ov{\overline}
\def\hlf{\frac{1}{2}}
\def\qrt{\frac{1}{4}}
\def\as{\alpha_s}
\def\at{\alpha_t}
\def\ab{\alpha_b}
\def\sq2{\sqrt{2}}
\newcommand{\smallz}{{\scriptscriptstyle Z}} %
\newcommand{\mz}{m_\smallz}
\newcommand{\smallw}{{\scriptscriptstyle W}}
\newcommand{\mw}{m_\smallw} 
\newcommand{\smallh}{{\scriptscriptstyle H}}
\newcommand{\mh}{m_\smallh}
\newcommand{\mt}{m_t}
\newcommand{\wh}{w_\smallh}
\def\th{t_\smallh}
\def\zh{z_\smallh}
\newcommand{\text}[1]{#1}
\newcommand{\Mvariable}[1]{#1}
\newcommand{\Mfunction}[1]{#1}
\newcommand{\Muserfunction}[1]{#1}

\newenvironment{appendletterA}
 {
  \typeout{ Starting Appendix \thesection }
  \setcounter{section}{0}
  \setcounter{equation}{0}
  \renewcommand{\theequation}{A\arabic{equation}}
 }{
  \typeout{Appendix done}
 }
\newenvironment{appendletterB}
 {
  \typeout{ Starting Appendix \thesection }
  \setcounter{equation}{0}
  \renewcommand{\theequation}{B\arabic{equation}}
 }{
  \typeout{Appendix done}
 }

%
%
%
%
%

\begin{titlepage}
\nopagebreak
{\flushright{
        \begin{minipage}{5cm}
         IFIC/07-55 \\
         RM3-TH/07-12 \\
         IFUM-902/FT \\ 
        {\tt hep-ph/yymmnnn}
        \end{minipage}        }

}
\renewcommand{\thefootnote}{\fnsymbol{footnote}}
\vskip 2.5cm
\begin{center}
\boldmath
{\Large\bf Scalar Particle Contribution  \\[7pt]
to Higgs Production via Gluon Fusion at NLO}\unboldmath
\vskip 1.0cm
{\large  R.~Bonciani\footnote{Email:
Roberto.Bonciani@ific.uv.es}},
\vskip .2cm
{\it Departamento de F\'{\i}sica Te\`orica, 
IFIC, CSIC -- Universidad de 
Valencia, \\
E-46071 Valencia, Spain}
\vskip .2cm
{\large G.~Degrassi\footnote{Email: degrassi@fis.uniroma3.it}},
\vskip .2cm
{\it Dipartimento di Fisica, Universit\`a di Roma Tre and 
INFN, Sezione di Roma Tre, \\ Via della Vasca Navale~84, I-00146 Rome, Italy} 
\vskip .2cm
{\large A.~Vicini\footnote{Email: Alessandro.Vicini@mi.infn.it}}
\vskip .2cm
{\it Dipartimento di Fisica, Universit\`a di Milano and
INFN, Sezione di Milano, \\
Via Celoria 16, I--20133 Milano, Italy} 
\end{center}
\vskip 0.7cm

\begin{abstract}
We consider the gluon fusion  production cross section of a scalar Higgs boson
in models where  fermion and scalar massive colored 
particles are present.  We report analytic expressions for the matrix elements 
of $gg\to Hg$, $q\bar{q}\to Hg$, and $qg\to Hq$ processes completing
the calculation of the NLO QCD corrections in these extended scenarios.
The formulas are written in a complete general case, 
allowing a flexible use for different theoretical models. 
Applications of our results to two different models are presented: i)
a model in which the SM Higgs sector is augmented by a weak doublet scalar
in the $SU(N_c)$ adjoint representation. ii)  The MSSM, 
in the limit of  neglecting the gluino contribution 
to the cross section. 

\vskip .4cm
{\it Key words}: Feynman diagrams, Multi-loop calculations, Higgs physics

{\it PACS}: 11.15.Bt; 12.38.Bx; 13.85.Lg; 14.80.Bn; 14.80.Cp. 
\end{abstract}
\vfill
\end{titlepage}    
\setcounter{footnote}{0}
%

\section{Introduction}

The Higgs mechanism for the electroweak symmetry breaking is the still untested
part of the Standard Model (SM). The search for the Higgs boson is one of 
the most important goals of the present experimental program at the Tevatron 
and, in the near future, at the Large Hadron Collider (LHC) at CERN.

Electroweak (EW) precision physics data and the direct search limit from LEP
constrain the possible values of the Higgs mass in the SM quite strongly, 
with a solid indication that a SM Higgs boson should be lighter
than 200 GeV. However, in many extentions of the SM the above bound does not 
apply because of the presence of new physics (NP) that can affect the EW fit 
allowing a higher value for the Higgs mass. 
Theoretical arguments based on perturbative unitarity, triviality and 
fine-tuning indicate that  the   crucial mass range to be investigated
is up to the TeV scale. The search of 
the Higgs boson at the LHC has, therefore, to be supported by an accurate 
theoretical knowledge of the production cross sections, the decay modes,
and the important background processes in this range (for a general 
review see Ref.~\cite{Dj}).

Due to the gluon luminosity, the main production mechanism for a scalar Higgs 
boson at the LHC is the gluon fusion process 
($pp \to H+X$) \cite{H2gQCD0}. 
Its cross section in the SM  is, all over the range of interesting values 
of the Higgs mass, one order of magnitude bigger than that of
the  other main production mechanisms, the Vector Boson Fusion (VBF) \cite{VBF},
and the production in association with heavy quarks and vector bosons
\cite{QQT}.

Differently from the other production mechanisms, the gluon fusion is a 
process that 
starts at $\mathcal{O}(\alpha_S^2 G_{\mu})$, i.e. at the one-loop level. 
In fact, the Higgs boson does not couple to gluons directly, but only via a 
loop of  colored particles. Thus, the gluon fusion process is
the Higgs production mechanism where NP can play the most relevant role
changing significantly the value of the production cross section.

The first predictions for the gluon fusion Higgs boson production cross 
section in the SM  dates in the late seventies \cite{H2gQCD0}. More than
fifteen years later, in 1991, the calculation of the NLO QCD corrections was
completed in the infinite top mass, $m_t$,  limit \cite{H2gQCD1} and, 
successively, retaining the full dependence on the mass of the heavy fermion 
that runs in the loops \cite{QCDg2}. The total effect of the NLO QCD corrections
is the increase of the LO cross section by a factor 1.5--1.7, giving a residual 
renormalization/factorization scale dependence of about 30\%. The unexpected 
size of the NLO QCD radiative corrections, that seem to spoil the validity of
perturbation theory, motivated, at the beginning of 2000, the calculation of 
the NNLO QCD corrections, performed in the infinite  $m_t$ limit \cite{H2gQCD2}.
The calculation shows a good convergence of the perturbative series. The NNLO 
corrections are sizable, but, nevertheless, smaller that the NLO ones. Moreover,
the QCD bands of variation of the renormalization/factorization scale overlap
with the ones of the NLO calculation. The NNLO corrections
enhance the cross section of an additional 15\% (of the NLO
results). Moreover, they improve the stability against 
renormalization/factorization scale variations.
Furthermore, the effect due to the resummation of soft-gluon radiation at the
NNLL accuracy has been evaluated in Ref.~\cite{bd4}. Besides an additional
enhancement of the cross section of the order of some percents (up to 6\%), 
the effect mainly lies in a strong reduction of the scale dependence.
The remaining theoretical uncertainty, due to higher-order QCD corrections, 
has been estimated to be smaller than 10\%. 
This estimate was confirmed recently by the NNNLO calculation of
Ref.~\cite{Moch:2005ky}.

Because of the high accuracy reached in the evaluation of the QCD
corrections, also the EW  NLO corrections to gluon fusion were
recently taken into account. In Ref.~\cite{DjG}, they were evaluated in
the infinite $m_t$ limit, giving a correction of less than 1\%. In
Ref.~\cite{ABDV}, the contributions coming from Feynman diagrams with a
closed loop of light fermions were calculated in a closed analytic
form, expressing the formulas in terms of generalized harmonic
polylogarithms (GHPLs) \cite{AB2}. It turned out that they are sizeable. In
particular, in the intermediate Higgs mass range, from 114 GeV up the
the $2\, m_W$ threshold, these corrections increase the LO cross
section by an amount of 4 to 9\%. For $m_H > 2 \,m_W$, they change sign
and reduce the LO cross section; however, in this region the
light-fermion corrections are quite small, reaching at most a -2\%.
In Ref.~\cite{DM1}, also the remaining EW corrections due to
the top quark were calculated as a Taylor expansion in
$m_H^2/(4m_W^2)$. This result is valid in the  $m_H \leq 2\, m_W$ range in which
the corrections due to the top quark have opposite sign with respect to 
the light-fermion contribution. However, the former are smaller
in size, reaching at most a 15\% of the latter.

Several efforts have also been devoted to the calculation of radiative 
corrections  (mainly QCD) to less inclusive quantities, like the
transverse momentum ($q_T$) distribution 
\cite{ehsv,BaurGlover,QTnlo,BCFG,Bozzi:2007pn} and the rapidity distribution  
\cite{Babis1,Babis2,Bozzi:2007pn}. 
All these results have been implemented in two Monte Carlos that 
calculate fully-differential distributions at the NNLO 
 \cite{Babis3,Catani:2007vq}.

The Higgs boson  gluon fusion production  cross section
was also  extensively studied in the Minimal Supersymmetric Standard Model 
(MSSM). The effects of the squarks  on the  production cross section for the 
neutral CP-even $h$ and $H$  Higgs bosons  were considered 
in Ref.~\cite{Dawson:1996xz}, including the NLO QCD corrections
evaluated in the  heavy squark mass limit. 
This approximation has been relaxed in Ref.~\cite{Muhlleitner:2006wx} 
where the full dependence on the squarks masses has been retained.
The complete MSSM NLO QCD corrections to the Higgs bosons production, 
in the heavy SUSY particles limit, has been presented in Ref.~\cite{Harlander:2003bb}. In general, these corrections lead to a NLO $K$ 
factor that differs from the corresponding SM one by  an amount  less 
than 5\%, with the exception of regions where the squark and quark 
contributions interfere negatively  giving rise to a  MSSM production 
cross section much smaller than the SM one \cite{Djouadi:1998az}.
Not inclusive quantities were also studied in the MSMM
\cite{MSSMH+j}.

As already pointed out, the Higgs boson production via gluon fusion,
as well as the Higgs decay into two photons, are processes sensitive to any kind
of NP. Thus, it would be desirable to have predictions for these quantities,
at the level of NLO QCD corrections, provided in full generality, 
i.e. not constrained  by a particular theoretical model but flexible enough 
to be used for different models, and, if possible, expressed in an 
analytic form easy to be evaluated numerically.

The aim of this paper is to analyze in a, as much as possible, model 
independent way the contribution of colored scalar particles to the
gluon fusion Higgs boson production cross section at the NLO level in the QCD 
corrections.  In this spirit, we present here general analytic formulas for 
the NLO 
corrections to the production cross section of a Higgs boson via 
gluon fusion, $\sigma(pp \to H+X)$. In particular, we provide analytic
expressions for the NLO QCD corrections to the partonic processes 
$gg\to Hg$, $q\bar{q}\to Hg$, and $qg\to Hq$, in the two cases in which 
a  fermion or a  scalar run in the loops. Together with the
analytic results of Refs.~\cite{Aglietti:2006tp,Anastasiou:2006hc}, 
where the two-loop virtual QCD correction to the gluon fusion process 
where evaluated, the present work completes the NLO calculation of the 
Higgs production cross section in the presence of colored scalar particles.

The paper is organized as follows. In Section \ref{sec3}, we provide all the 
analytic formulas and we discuss the
validity of the heavy fermion/scalar mass limit. Applications of our results
to two different models are presented in the following section.
In Section \ref{MW} we consider a model, proposed by Manohar and Wise 
\cite{Manohar:2006gz}, 
in which the Standard Model is supplemented by an additional scalar colored 
isospin doublet. We discuss, at the NLO level, the effects of these
additional scalar particles on  the production cross section of the standard 
Higgs boson and on its decay width  $H \to \gamma \gamma$.
In Section \ref{MSSM}, we consider the contributions due to the squarks in the 
MSSM. We focus on the NLO QCD corrections 
(exchange of gluons), neglecting the contributions coming from the
gluino.
Finally, in Section \ref{CONC} we present our conclusions.
%

\section{Higgs  Production via Gluon Fusion at NLO \label{sec3}}

In this section we present analytic results for the NLO QCD corrections to  
Higgs boson production via the gluon fusion mechanism. 
Being the Higgs boson neutral under $SU(N_c)$, its coupling to the gluons is
mediated by a loop of colored particles. To discuss the gluon fusion mechanism
in a general way   we assume as colored particles one  fermion and one scalar 
in a generic $R_{1/2},\, R_0$ $SU(N_c)$ representation, respectively. The 
extension to more fermions or scalars  is trivial. The coupling's strengths
of these particles to the Higgs are assumed to be:
\be
HFF= g \,\lambda_{1/2}\, \frac{m_{1/2}}{2\,\mw},
~~~~~~~~~~~HSS = g \,\lambda_{0}\, \frac{A^2}{\mw}~,
\ee
where $g$ is the $SU(2)$ coupling, $\mw$ is the W mass, $m_{1/2}$ is the
fermion mass, $A$ is a generic coupling
with the dimension of mass and
$\lambda_i$ are numerical coefficients\footnote{The SM is recovered with
$ \lambda_{1/2}=1, \: \lambda_0 = 0$, $N_c=3$ and
$R_{1/2} = {\mathbf 3}$.}.

The hadronic cross section for the Higgs production via gluon fusion at 
center-of-mass energy $\sqrt{s}$, can be 
written as:
\bea
\sigma(h_1 + h_2 \to H+X) & = & 
          \sum_{a,b}\int_0^1 dx_1 dx_2 \,\,f_{a,h_1}(x_1,\mu_F^2)\,
         f_{b,h_2}(x_2,\mu_F^2) \times \nonumber\\
& & \times
\int_0^1 dz~ \delta \left(z-\frac{\tau_H}{x_1 x_2} \right)
\hat\sigma_{ab}(z) \, ,
\label{sigmafull}
\eea
where $\tau_H= \mh^2/s$, $\mu_F$ is the factorization scale,
$f_{a,h_i}(x,\mu_F^2)$, the parton density of the colliding hadron $h_i$
for the parton of type $a, \,(a = g,q,\bar{q})$ and $\hat\sigma_{ab}$ the 
cross section for the partonic subprocess $ ab \to H +X$ at the center-of-mass 
energy  $\hat{s}=x_1 x_2 s=\mh^2/z$. The latter can be written as:
\be
\hat\sigma_{ab}(z)=
\sigma^{(0)}\,z \, G_{ab}(z) \, ,
\label{Geq}
\ee
where 
\be
\sigma^{(0)}  =  
\frac{G_\mu \alpha_s^2 (\mu_R^2)}{128\, \sqrt{2} \, \pi}
\left|   \sum_{i=0,1/2} \lambda_i \left(\frac{A^2}{m_0^2} \right)^{1-2 i}\,
T(R_i)\, {\mathcal G}^{(1l)}_{i}
       \right|^2 
\label{ggh}
\ee
is the Born-level contribution  with $m_0$  the mass of the scalar particle,
\bea
{\mathcal G}^{(1l)}_{1/2} & = & - 4 y_{1/2}
 \left[ 2 - \left( 1 -4 y_{1/2} \right)  \, H(0,0,x_{1/2}) \right] \, ,
\label{eq:3} \\
{\mathcal G}^{(1l)}_{0} & = & 4 y_0 \left[ 1 + 2 \, y_0\,
                 H(0,0,x_{0}) \right]
\label{eq:4}
\eea
and $T(R_i)$ is the matrix normalization factor of the $R_i$ representation
($T(R) =1/2$ for the fundamental  representation of $SU(N_c)$, 
$T(R) =N_c$ for the adjoint one). In Eqs.~({\ref{eq:3}-\ref{eq:4})
\be
y_i \equiv \frac{m_i^2}{\mh^2}, ~~~~~~~~
x_i \equiv \frac{\sqrt{1- 4 y_i} - 1}{\sqrt{1- 4 y_i} + 1}~~~~~~~~
i=0,1/2  \, ,
\label{defx}
\ee
and, employing the standard notation for the Harmonic Polylogarithms (HPLs),  
$H(0,0,z)$ labels a HPL of weight 2
that results to be\footnote{All the analytic  continuations are obtained with
the replacement $-\mh^2 \to -\mh^2 - i \epsilon$}
\be
H (0,0,z ) = \frac{1}{2} \log^2 (z)~.
\ee

Up to NLO terms, we can write
\be
G_{a b}(z)  =  G_{a b}^{(0)}(z) 
          + \frac{\alpha_s (\mu^2_R)}{\pi} \, G_{a b}^{(1)}(z) \, ,
\ee
with
\bea
G_{a b}^{(0)}(z) & = & \delta(1-z) \,\delta_{ag}\, \delta_{bg} \, , \\
G_{g g}^{(1)}(z) & = & \delta(1-z) \left[
            C_A \, \frac{\pi^2}3  
	  + \beta_0 \ln \left( \frac{\mu_R^2}{\mu_F^2} \right)
          + \sum_{i=0,1/2} {\mathcal G}^{(2l)}_{i} \right]  \nn \\
& & + P_{gg} (z)\ln \left( \frac{\hat{s}}{\mu_F^2}\right) +
    C_A\, \frac4z \,(1-z+z^2)^2 \,{\cal D}_1(z) +  C_A\, {\cal R}_{gg}  \, , 
\label{real} \\
G_{q \bar{q}}^{(1)}(z) & = &   {\cal R}_{q \bar{q}} \, , \label{qqbar}\\
G_{q g}^{(1)}(z) & = &  P_{gq}(z) \left[ \ln(1-z) + 
 \frac12 \ln \left( \frac{\hat{s}}{\mu_F^2}\right) \right] + {\cal R}_{qg} \, .
\label{qg}
\eea
We discuss the various NLO contributions. 

i) The $gg$ channel (Eq.~(\ref{real}))
involves virtual and real corrections. The former, regularized by the infrared 
singular part of the cross section  $ gg \to H g$, are displayed in the
first row of Eq.~(\ref{real}) where  $\beta_0 = (11\, C_A - 4\, n_f \,T(R_f) -
n_s\, T(R_s))/6$  with $n_f\: (n_s)$  the  number of active fermion (scalar) 
flavor in the  representation $R_f\: (R_s)$. The functions 
$ {\mathcal G}^{(2l)}_{i}$ containing the mass-dependent contribution of 
the  two-loop  virtual  corrections, can
be  cast in the following form:
\bea
{\mathcal G}^{(2l)}_{i} & = &
        \lambda_i \left(\frac{A^2}{m_0^2} \right)^{1-2 i}\, T(R_i)
               \Biggl( C(R_i)\, {\mathcal G}^{(2l, C_R)}_{i} (x_i)+
                    C_A \,{\mathcal G}^{(2l, C_{A})}_{i} (x_i)  \Biggr) \nn \\
& & \times
     \left( \sum_{j=0,1/2}  \lambda_j \left(\frac{A^2}{m_0^2} \right)^{1-2 j}\,
       T(R_j)\, {\mathcal G}^{(1l)}_{j} \right)^{-1} + h.c.
\label{G2}
\eea
where $C(R_i)$ is the Casimir factor of the $R_i$ representation 
(in particular,
for the fundamental and the adjoint representations of $SU(N_c)$ we have
$C_F = (N_c^2-1)/(2N_c)$ and $C_A = N_c$, respectively).
Explicit analytic expressions for ${\mathcal G}^{(2l)}_{i}\: (i=0,1/2)$ 
given in terms of HPLs can be found  in Ref.~\cite{Aglietti:2006tp}.
It should be noticed that ${\mathcal G}^{(2l,C_R)}_{i}$ depend upon 
the choice of the renormalized parameters (masses and couplings). In 
Ref.~\cite{Aglietti:2006tp} expressions for ${\mathcal G}^{(2l,C_R)}_{i}$ 
with $\overline{\mbox{MS}}$ or on-shell (OS) parameters are presented. In the 
case of single heavy fermion, i.e. $\lambda_0=0$, 
${\mathcal G}^{(2l)}_{1/2}$ become independent of
the renormalized mass chosen and goes to the well know result 
$- 3/2 \,C(R_{1/2}) + 5/2\, C_A$, that can be also obtained via an effective 
theory calculation \cite{H2gQCD1}. The case of a single heavy 
scalar ($\lambda_{1/2} =0$) is actually more complicated because,
in general, the coupling of a colored scalar particle to the Higgs boson
is not directly proportional to the mass of the scalar and therefore different
renormalization prescriptions for $A$ and $m_0$ can be employed. In case
an $\overline{\mbox{MS}}$ prescription is employed both for $A$ and $m_0$, the 
function ${\mathcal G}^{(2l)}_{0}$ tends, for large values of $m_0$, to the 
constant  value $9/2 \,C(R_{0}) +  C_A$ independent on the  
$\overline{\mbox{MS}}$  subtraction scale.

The second row of Eq.~(\ref{real}) contains the non-singular contribution
from the real gluon emission in the gluon fusion process where 
\be
{\cal D}_i (z) =  \left[ \frac{\ln^i (1-z)}{1-z} \right]_+  \label {Dfun}
\ee
are the plus distributions and
\be
P_{gg} (z)= 2\,  C_A\,\left[ {\cal D}_0(z) +\frac1z -2 + z(1-z) \right]
\label{Pgg}
\ee
is the LO Altarelli-Parisi splitting function. The function 
${\cal R}_{gg}$ can be written as
\be
{\cal R}_{gg} = \frac1{z(1-z)}\int_0^1 \frac{d v}{v (1-v)} \left\{
 \frac{8\,z^4 \left| {\cal A}_{gg}(\hat{s},\hat{t},\hat{u})\right|^2 }{  
\left| \sum_{j=0,1/2}  \lambda_j \left(\frac{A^2}{m_0^2} \right)^{1-2 j}\,
       T(R_j)\, {\mathcal G}^{(1l)}_{j} \right|^2 }  - (1-z+z^2)^2 \right\} ,
\ee
where $\hat{t} = -\hat{s} (1-z) (1-v),\,\hat{u} = -\hat{s} (1-z) v$,
with 
\be
\left| {\cal A}_{gg}(s,t,u) \right|^2  =
 |A_2 (s,t,u)|^2 + |A_2 (u,s,t)|^2 + |A_2 (t,u,s)|^2 +
      |A_4 (s,t,u)|^2 .
\label{Agg}
\ee
Furthermore, the functions $A_2$ and $A_4$ can be cast in the following form:
\bea
A_2 (s,t,u) & = & \sum_{i=0,1/2}  
 \lambda_i \left(\frac{A^2}{m_0^2} \right)^{1-2 i}\,
       T(R_i)\,y_i^2 \,\left[ b_i (s_i,t_i,u_i) + b_i (s_i,u_i,t_i) \right] , \\
A_4 (s,t,u) & = & \sum_{i=0,1/2}  
 \lambda_i \left(\frac{A^2}{m_0^2} \right)^{1-2 i}\,
       T(R_i)\,y_i^2 \,\left[ c_i (s_i,t_i,u_i) + c_i (t_i,u_i,s_i) 
                   \right. \nn \\
& &~~~~~~~~~~~~~~~~~~~~~~~~~~~~~~~~~~~
            \left.  +\,  c_i (u_i,s_i,t_i) \right] ,
\label{A24fun}
\eea
with 
\be
s_i \equiv \frac{s}{m_i^2},~~~~~t_i \equiv \frac{t}{m_i^2}, ~~~~~~~
u_i \equiv \frac{u}{m_i^2}.
\ee
We find
\bea
b_{1/2}(s,t,u) &=&  B_{1/2} (s,t,u)
+ \frac{s}4 \left[H(0,0,x_{1/2})-H(0,0,x_s) \right] 
\nn \\
& & - \left(\frac{s}{2} 
-\frac{s^2}{s+u}\right) \left[H(0,0,x_{1/2})-H(0,0,x_t)\right] \nn \\
& & - \frac{s}{8}  \text{H_3}(s,u,t)+
\frac{s}4\,  \text{H_3}(t,s,u) ,
\label{eq:b12}\\
b_{0}(s,t,u) &=& -\frac12 B_0(s,t,u) , \label{eq:b0}
\eea 
\bea
\! \! \! \! c_{1/2}(s,t,u) & = & C_{1/2} (s,t,u)
+\frac{1}{2 \,y_{1/2}} 
\left[H(0,0,x_{1/2})-H(0,0,x_s)\right] 
+ \frac1{4\,y_{1/2}} \text{H_3}(s,u,t) , 
\label{eq:c12}\\
\! \! \! \! c_{0}(s,t,u) & = & -\frac12 C_0(s,t,u) ,
\label{eq:c0}
\eea
where
$$
x_a \equiv \frac{\sqrt{1- 4/a} - 1}{\sqrt{1- 4/a} + 1}~~~~~~~~~~~~(a=s,t,u) 
$$
and
\bea  
B_i(s,t,u) &=& \frac{s (t-s)}{s+t}+
\frac{2 \left(t u^2+2 s t u\right)}{(s+u)^2}
\left[\sqrt{1-4 y_i}\, H(0,x_i)-
\sqrt{1-4/t}\, H(0,x_t) \right] \nn \\
& & - \left( 1+ \frac{t u }{s} \right) H(0,0,x_i) +H(0,0,x_s)
\nn \\
& & - 2\left(\frac{2s^2}{(s+u)^2}-1-\frac{ t u}{s}\right) 
\left[H(0,0,x_i)-H(0,0,x_t) \right] \nn \\
& & + \frac{1}{2} \left(\frac{t u}{s} 
+ 3\right) \text{H_3}(s,u,t)-  \text{H_3}(t,s,u) ,
\label{eq:Bi} \\
C_i(s,t,u) &=&  -2 s - 
2 \left[ H(0,0,x_i)- H(0,0,x_s)\right]
- \text{H_3}(u,s,t) .
\label{eq:Ci}
\eea
In Eqs.~(\ref{eq:b12}--\ref{eq:Ci}) $H(0,x) \equiv \ln(x)$, and
the function $H_3$ is symmetric under the interchange of its last two
arguments, i.e. $H_3(a,b,c) = H_3(a,c,b)$,
as can be seen from its integral representation:
\bea
H_3(a,b,c) &=& \int_0^1 d x \frac1{x(1-x) + a/(b\,c)} \left\{
       \ln [1 - b x(1-x)] + \ln [1- c x(1-x)]~~~~~ \right. \nn \\
&& ~~~~~~~~~~~~~~~~~~~~~~~~~~~~~~~~ \left. - \ln [ 1 -(a+b+c) x(1-x)] \right\}.
\eea
An explicit  analytic expression for $H_3(a,b,c)$ can be found, for instance, 
in Ref.~\cite{ehsv}. Using the notations of \cite{ehsv}, we have
$H_3(a,b,c)=-W_3(b,a,c,\,a+b+c)$\footnote{In Ref.~\cite{ehsv}, in  the function 
$W_3$  the mass of the heavy particle is explicitely written in
the integral representation. Instead the $H_3$ function has as input parameters 
``reduced'' variables, i.e. Mandelstam variables divided by the heavy particle 
mass. Therefore, in order to have the correct formal expression for $H_3$ 
one has  to put in the formulas of Ref.~\cite{ehsv} $m_f=1$ and consider the 
variables $s$, $t$ and $u$ as reduced variables.}.

In the case of a single heavy fermion or a single heavy scalar 
or both fermion and scalar heavy ${\cal R}_{gg} \to -11 (1-z)^3/(6 z)$.

\vspace{0.3cm}
ii) The $ q \bar q \to H g $ annihilation channel, Eq.~(\ref{qqbar}),
can be written as
\be
{\cal R}_{q \bar q} = \frac{128}{27} 
 \frac{z\,(1-z)\, \left| {\cal A}_{q \bar q}(\hat{s},\hat{t},\hat{u})\right|^2}
{\left| \sum_{j=0,1/2}  \lambda_j \left(\frac{A^2}{m_0^2} \right)^{1-2 j}\,
       T(R_j)\, {\mathcal G}^{(1l)}_{j} \right|^2 }  \, ,
\ee
with
\be
{\cal A}_{q \bar q} ( s,t,u)  =  \sum_{i=0,1/2}  
 \lambda_i \left(\frac{A^2}{m_0^2} \right)^{1-2 i}\,
       T(R_i)\,y_i \, d_i (s_i,t_i,u_i)~.  
\ee
We find
\bea
d_{1/2}(s,t,u) &=&  D_{1/2} (s,t,u)
- 2 \left[H(0,0,x_{1/2})- H(0,0,x_s) \right] ,
\label{eq:d12}\\
d_{0}(s,t,u) &=& -\frac12 D_0(s,t,u) , \label{eq:d0}
\eea 
with
\bea
D_i(s,t,u) &=& 4 +  \frac{4\, s}{(t+u)}
\left[\sqrt{1-4 y_i}\, H(0,x_i)-
\sqrt{1-4/s}\, H(0,x_s) \right] \nn \\
& & + \frac8{t+ u } \left[ H(0,0,x_i) - H(0,0,x_s) \right] .
\label{di}
\eea
In the case of a single heavy fermion or a single heavy scalar 
or both fermion and scalar heavy ${\cal R}_{q \bar q} \to 32 (1-z)^3/(27 z)$.

\vspace{0.3cm}
iii) Finally we consider the quark-gluon scattering channel,
$q g \to q H $. In Eq.~(\ref{qg})
$P_{gq}$ is the LO Altarelli-Parisi splitting function
\be
P_{gq} (z) = C_F \,\frac{1 + (1-z)^2}z \, ,
\ee
while the function  ${\cal R}_{qg}$ can be written as
\bea
{\cal R}_{qg} \! \! & = & \! \! C_F \! \int_0^1 \! \frac{d v}{(1-v)} \left\{
\frac{ 1 + (1-z)^2 v^2}{[1-(1-z) v]^2} 
\frac{2 \,z \left| {\cal A}_{qg}(\hat{s},\hat{t},\hat{u})\right|^2 }{  
\left| \sum_{j=0,1/2}  \lambda_j \left(\frac{A^2}{m_0^2} \right)^{1-2 j} \!
       T(R_j)\, {\mathcal G}^{(1l)}_{j} \right|^2 }  
   - \frac{1+(1 \! - \! z)^2}{2 z} \right\} \nn \\
& & + \frac12 C_F z \, ,
\eea
where
\be
{\cal A}_{qg}(\hat{s},\hat{t},\hat{u}) = 
{\cal A}_{qq}(\hat{t},\hat{s},\hat{u}) .
\ee
In the case of a single heavy fermion or a single heavy scalar 
or both fermion and scalar heavy  ${\cal R}_{q g} \to 2\,z/3  - (1-z)^2/z$.

The expressions reported above are, for the fermionic case, in full
agreement with the known results in the literature
\cite{ehsv,BaurGlover}.  They give the exact, i.e. for any value of
the particle masses, NLO contribution and correspondingly the exact
K-factor defined as the ratio between the NLO and LO cross sections.
It is interesting to compare the value of the exact K-factor with the
one that can be obtained via an improved effective theory
calculation. By the latter we mean a result in which the effective NLO
cross section is obtained by multiplying the exact LO cross section by
the ${\cal O}(\as)$ corrections evaluated in the heavy particle limit
\cite{Kramer:1996iq}. As discussed above, while for fermions the NLO
contribution in the limit of heavy mass is independent upon the
definition of the renormalized mass used, the heavy scalar NLO
contribution is actually dependent on the renormalization conditions
chosen.
%
\begin{figure}
\bc
\begin{picture}(0,0)%
\includegraphics{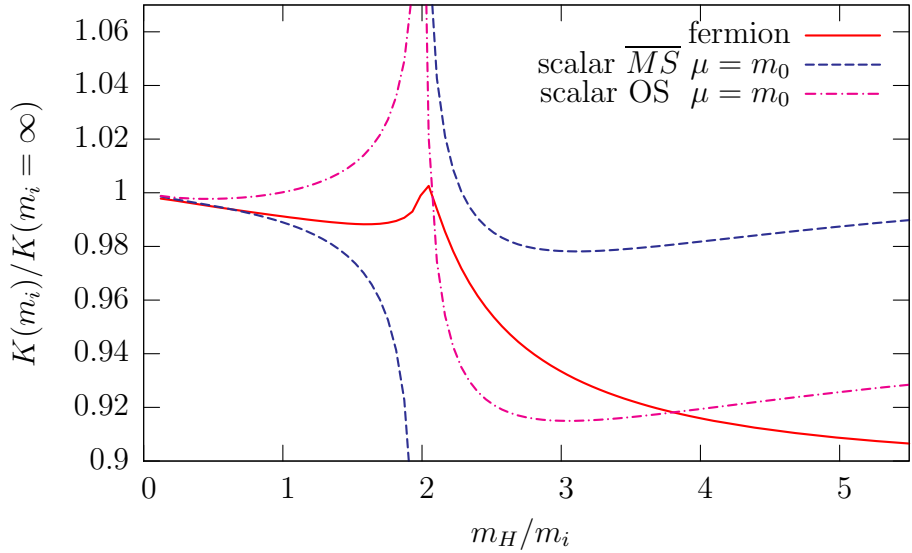}%
\end{picture}%
\begingroup
\setlength{\unitlength}{0.0200bp}%
\begin{picture}(18000,10800)(0,0)%
\put(2475,1650){\makebox(0,0)[r]{\strut{} 0.9}}%
\put(2475,2662){\makebox(0,0)[r]{\strut{} 0.92}}%
\put(2475,3674){\makebox(0,0)[r]{\strut{} 0.94}}%
\put(2475,4685){\makebox(0,0)[r]{\strut{} 0.96}}%
\put(2475,5697){\makebox(0,0)[r]{\strut{} 0.98}}%
\put(2475,6709){\makebox(0,0)[r]{\strut{} 1}}%
\put(2475,7721){\makebox(0,0)[r]{\strut{} 1.02}}%
\put(2475,8732){\makebox(0,0)[r]{\strut{} 1.04}}%
\put(2475,9744){\makebox(0,0)[r]{\strut{} 1.06}}%
\put(2750,1100){\makebox(0,0){\strut{} 0}}%
\put(5373,1100){\makebox(0,0){\strut{} 1}}%
\put(7995,1100){\makebox(0,0){\strut{} 2}}%
\put(10618,1100){\makebox(0,0){\strut{} 3}}%
\put(13241,1100){\makebox(0,0){\strut{} 4}}%
\put(15864,1100){\makebox(0,0){\strut{} 5}}%
\put(550,5950){\rotatebox{90}{\makebox(0,0){\strut{}$K(m_i)/K(m_i=\infty)$}}}%
\put(9962,275){\makebox(0,0){\strut{}$m_H/m_i$ }}%
\put(14950,9675){\makebox(0,0)[r]{\strut{}fermion}}%
\put(14950,9125){\makebox(0,0)[r]{\strut{}scalar $\overline{MS}~\mu=m_0 $}}%
\put(14950,8575){\makebox(0,0)[r]{\strut{}scalar OS~~$\mu=m_0 $}}%
\end{picture}%
\endgroup
\caption{\label{fig1} Exact K-factor normalized to the effective one.
The different lines represent a single heavy fermion (solid line) and a 
single heavy scalar whose mass is renormalized in the $\overline{\mbox{MS}}$ 
(dashed line) or in the on-shell scheme (dashed-dotted line).}
\ec
\end{figure}

In Fig.~\ref{fig1} we plot the exact NLO K-factor normalized to the effective
one as a function of $\mh /m_i$, for the case of a single fermion 
(continuos line)  and of a single scalar. 
For the latter we consider two options: 
i) On-shell condition for $m_0$ and $\overline{\mbox{MS}}$ renormalization
of the coupling $A$ defined at the $\overline{\mbox{MS}}$ $\mu$ scale
$\mu = m_0$, with 
$A(m_0)= m_0$ (dash-dotted line). ii) $\overline{\mbox{MS}}$
renormalization both 
for $A$ and $m_0$ with $\mu = m_0$  and $A(m_0)=m_0 (\mu)$ (dashed line).  
The results presented in the figure have been obtained
assuming a hadronic center-of-mass energy
$\sqrt{s}=14$ TeV, using the parametrization CTEQ6M \cite{cteq}
to describe the partonic content of the proton, and setting the factorization 
and renormalization scales equal to the Higgs boson mass. These choices
will be used also in the figures of the following sections.
From Fig.~\ref{fig1} it appears that in the fermion case the
difference between the exact and the effective K-factor is at most $10\%$
and that already when $ \mh < 2 \,m_{1/2}$ the difference is below  $1 \%$.
Instead, in the scalar case the situation is more complicated. 
Both cases i) and ii)  show a spike  at the opening of the 
$\mh = 2 \,m_0$ threshold. This spike is due to logarithmic and square
root singularities present in the two-loop one-particle irreducible (1PI) 
virtual corrections, coming from diagrams in which a scalar self-energy diagram 
is inserted in a one-loop vertex diagram. When the mass of the scalar is
renormalized on-shell, the mass-counterterm diagrams  show a square root
singularity that actually cancels the similar one  coming from the 1PI diagrams,
leaving only an unphysical logarithmic singularity (see dash-dotted line in 
Fig.~\ref{fig1}) related to the inadequateness  of the standard mass 
renormalization procedure in case of unstable particles.
Instead, in the case an $\overline{\mbox{MS}}$ renormalization for the
mass of the scalar is 
employed, the cancellation of the square root singularity between 1PI diagrams
and mass-counterterm diagrams does not take place anymore. Then both the
square root and the logarithmic unphysical singularities are left (see dashed 
line in Fig.~\ref{fig1}). 
In the region away from the threshold, the figure shows a good convergence 
to 1 for light Higgs masses. For a heavy Higgs, instead, we note a certain
deviation of the effective theory from the exact one. Nevertheless, this
deviation remains quite limited in size, and it reaches at most 9\%.
%

\section{Scalar Particle Effects on the Higgs Production Cross Section
\label{SCA}}

In this section we discuss the effect of colored scalar particles on the 
Higgs production cross section via gluon fusion. We consider two cases:
i) a model in which the SM 
fields are supplemented by a weak doublet of colored scalars. ii) The squark 
contribution in the MSSM.

\subsection{The Manohar-Wise Model \label{MW}} 

The model proposed by Manohar and Wise (MW) \cite{Manohar:2006gz}
is an extension of the SM that includes additional colored scalar fields
which transform in the $({\bf 8},{\bf 2})_{1/2}$ representation of 
$SU(3)\times SU(2)\times U(1)$. The choice of these additional  scalar 
fields is dictated by requirement of natural suppression of flavor changing
neutral currents.  The additional colored scalar weak doublet
\be
S^a =
\left(
\begin{array}{c}
S^{a}_+\\
S_0^a
\end{array}
\right)
=
\left(
\begin{array}{c}
S^{a}_+\\
\frac{S_{0R}^a+i S_{OI}^a}{\sqrt{2}}
\end{array}
\right)
\ee
with $a=1\dots8$ an adjoint color index, contains an electrically  charged and 
two neutral real scalars. Denoting the standard 
$({\bf 1},{\bf 2})_{1/2}$ Higgs field  by $H$, 
the most general potential can be written as \cite{Manohar:2006gz}
\bea
V &=& \frac{\lambda}{4} \left(H^{\dagger i}H_i-\frac{v^2}{2}
\right)^2
~+~2m_S^2~Tr S^{\dagger i} S_i \nonumber \\
 &+&\lambda_1 H^{\dagger i} H_i Tr S^{\dagger j} S_j
+\lambda_2 H^{\dagger i} H_j Tr S^{\dagger j} S_i
+ \left(
\lambda_3 H^{\dagger i} H^{\dagger j} Tr S_i S_j
~+~h.c.\right) +\cdots
\label{potential}
\eea
where $S = S^aT^a$, the trace is over color and $SU(2)$ indices are 
explicitly shown. The ellipses represent tri- and quadrilinear 
interaction terms of the fields $S^a$ not relevant for our discussion.
The tree-level mass spectrum of the 
colored octet scalars is found to be:
\bea
m_{S_+}^2  &=& m_S^2 + \lambda_1~\frac{v^2}4 \, , \nonumber \\
m_{S_{0R}}^2 &=& m_S^2 + (\lambda_1+ \lambda_2 + 2 \lambda_3)~\frac{v^2}4 
\, ,\nonumber\\
m_{S_{0I}}^2 &=& m_S^2 + (\lambda_1+\lambda_2 - 2 \lambda_3)~\frac{v^2}4 \, .
\label{masses} 
\eea
and the coupling to the standard Higgs are:
\bea
H S^{a}_+ S^{b}_- &=& g ~\frac{\lambda_1}4 \frac {v^2}{\mw} \delta^{ab} 
\, , \nonumber \\
H S^{a}_{0R} S^{b}_{OR} &=& g  \frac{\lambda_1+\lambda_2 + 2 \lambda_3}{8}~ 
\frac {v^2}{\mw} \delta^{ab} \, , \nonumber\\
H S^{a}_{0I} S^{b}_{0I} &=& 
g \frac{\lambda_1+\lambda_2 - 2 \lambda_3}{8}~\frac {v^2}{\mw} \delta^{ab} ~.
\label{couplings} 
\eea

\begin{figure}
\begin{center}
\begin{picture}(0,0)%
\includegraphics{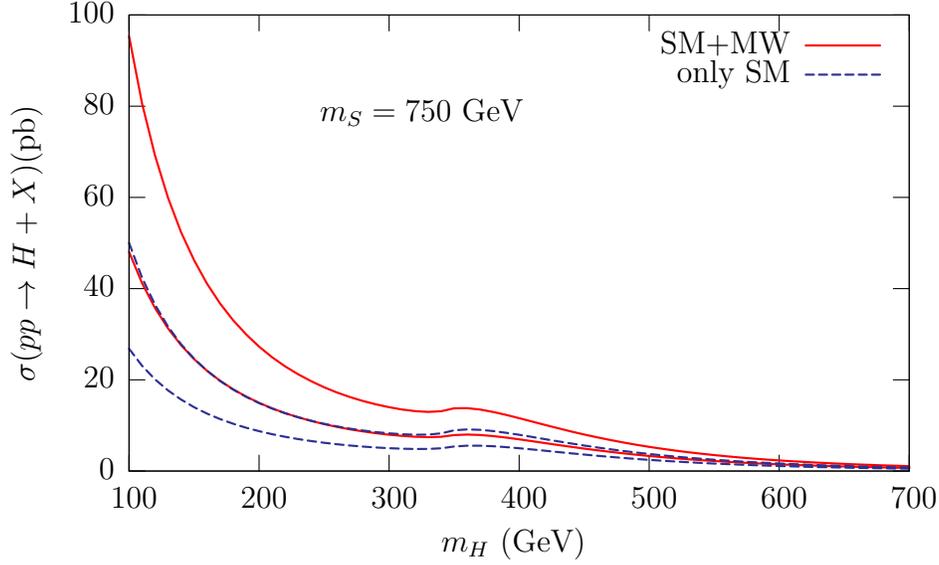}%
\end{picture}%
\begingroup
\setlength{\unitlength}{0.0200bp}%
\begin{picture}(18000,10800)(0,0)%
\put(2200,1650){\makebox(0,0)[r]{\strut{} 0}}%
\put(2200,3370){\makebox(0,0)[r]{\strut{} 20}}%
\put(2200,5090){\makebox(0,0)[r]{\strut{} 40}}%
\put(2200,6810){\makebox(0,0)[r]{\strut{} 60}}%
\put(2200,8530){\makebox(0,0)[r]{\strut{} 80}}%
\put(2200,10250){\makebox(0,0)[r]{\strut{} 100}}%
\put(2475,1100){\makebox(0,0){\strut{} 100}}%
\put(4925,1100){\makebox(0,0){\strut{} 200}}%
\put(7375,1100){\makebox(0,0){\strut{} 300}}%
\put(9825,1100){\makebox(0,0){\strut{} 400}}%
\put(12275,1100){\makebox(0,0){\strut{} 500}}%
\put(14725,1100){\makebox(0,0){\strut{} 600}}%
\put(17175,1100){\makebox(0,0){\strut{} 700}}%
\put(550,5950){\rotatebox{90}{\makebox(0,0){\strut{}$\sigma(pp\to H+X)({\rm pb})$}}}%
\put(9825,275){\makebox(0,0){\strut{}$m_H$~(GeV) }}%
\put(6075,8390){\makebox(0,0)[l]{\strut{}$m_S=750~{\rm GeV}$}}%
\put(14950,9675){\makebox(0,0)[r]{\strut{} SM+MW}}%
\put(14950,9125){\makebox(0,0)[r]{\strut{} only~SM}}%
\end{picture}%
\endgroup
\caption{\label{fig3} Higgs production cross sections as a function of the 
Higgs mass. The particles running in the loop are: only SM fermions (dashed
lines) or both SM fermions and MW scalars (solid lines). 
The lower curves in the two cases represent the LO cross section, while the
upper curves represent the NLO one. The couplings are
$\lambda_1 (m_S)=4,~\lambda_2 (m_S) =1,~\lambda_3 (m_S) =1/2 $.
The scalar mass is $m_S=750~{\rm GeV}$.  }
\end{center}
\end{figure}
Eqs.~(\ref{masses},\ref{couplings}) show that in this model $A/m_0 \sim v/m_S$ 
for $m_S \gg v$ ensuring the decoupling of the colored scalars as their mass 
increases.

To analyze the effect on the Higgs production cross section of this octet of 
colored scalars we plot, in Fig.~\ref{fig3}, the LO and NLO cross sections, 
as a function of $\mh$, including the scalar contribution and compare them 
with the SM results. The figure is drawn taking  $T(R_0)=C(R_0)=3$,
$m_S = 750$~GeV, 
$\lambda_1 (m_S) = 4, \, \lambda_2 (m_S)= 1, \, \lambda_3 (m_S)= 1/2,$
($\overline{\mbox{MS}}$ couplings  at the scale $\mu=m_S $)\footnote{This set
of parameters is consistent with the electroweak precision physics constraints
\cite{Manohar:2006gz}. In this model it is possible to generate a positive 
contribution to the $\rho$ parameter from the colored scalar sector allowing 
for  an heavier standard Higgs boson in the electroweak fit.}  and 
renormalizing the mass of the scalars on-shell. With this set of parameters
$A^2/m_0^2 \sim 0.1$, thus it acts as  a large suppression factor.
The figure shows that the NLO production cross section in this model is 
always significantly larger that the SM one. In particular, for small values 
of $\mh$ it is almost two times the SM cross section. 

The presence of an additional octet of scalar particle affects not only
the production cross section of the standard Higgs boson but also its decay
into two photons, that is a very relevant mode for Higgs searches up to
$~\mh= 140$~GeV. The formulas for the decay width $H \to \gamma \gamma$,
including the contribution of colored scalar particles evaluated at the NLO, 
can be found in Ref.~\cite{Aglietti:2006tp}. 
\begin{figure}
\bc
\begin{picture}(0,0)%
\includegraphics{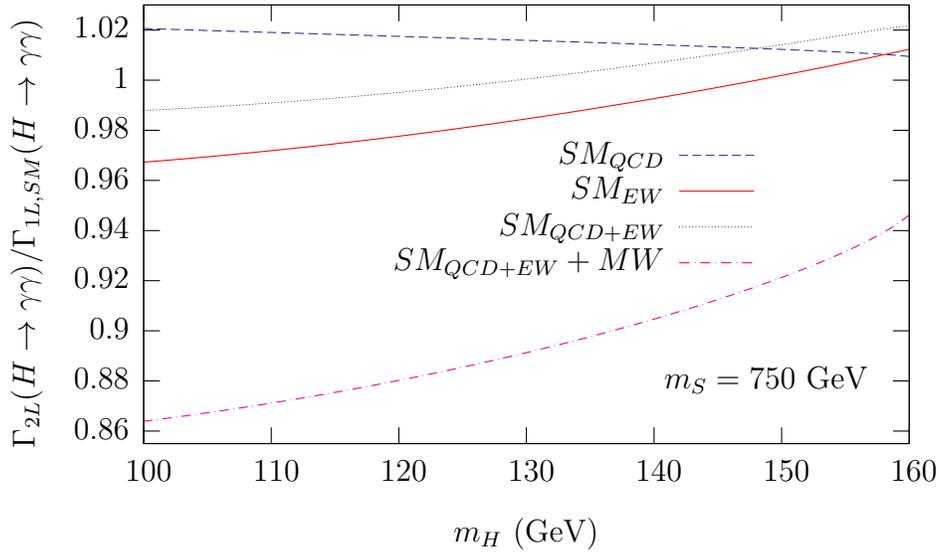}%
\end{picture}%
\begingroup
\setlength{\unitlength}{0.0200bp}%
\begin{picture}(18000,10800)(0,0)%
\put(2475,2215){\makebox(0,0)[r]{\strut{} 0.86}}%
\put(2475,3161){\makebox(0,0)[r]{\strut{} 0.88}}%
\put(2475,4106){\makebox(0,0)[r]{\strut{} 0.9}}%
\put(2475,5051){\makebox(0,0)[r]{\strut{} 0.92}}%
\put(2475,5996){\makebox(0,0)[r]{\strut{} 0.94}}%
\put(2475,6942){\makebox(0,0)[r]{\strut{} 0.96}}%
\put(2475,7887){\makebox(0,0)[r]{\strut{} 0.98}}%
\put(2475,8832){\makebox(0,0)[r]{\strut{} 1}}%
\put(2475,9777){\makebox(0,0)[r]{\strut{} 1.02}}%
\put(2750,1429){\makebox(0,0){\strut{} 100}}%
\put(5154,1429){\makebox(0,0){\strut{} 110}}%
\put(7558,1429){\makebox(0,0){\strut{} 120}}%
\put(9962,1429){\makebox(0,0){\strut{} 130}}%
\put(12367,1429){\makebox(0,0){\strut{} 140}}%
\put(14771,1429){\makebox(0,0){\strut{} 150}}%
\put(17175,1429){\makebox(0,0){\strut{} 160}}%
\put(550,5949){\rotatebox{90}{\makebox(0,0){\strut{}$\Gamma_{2L}(H\to\gamma\gamma)/\Gamma_{1L,SM}(H\to\gamma\gamma)$}}}%
\put(9962,275){\makebox(0,0){\strut{} $m_H$ (GeV) }}%
\put(12367,3161){\makebox(0,0)[l]{\strut{} $m_S=750$~GeV}}%
\put(12572,7414){\makebox(0,0)[r]{\strut{}$SM_{QCD}$}}%
\put(12572,6739){\makebox(0,0)[r]{\strut{}$SM_{EW}$}}%
\put(12572,6064){\makebox(0,0)[r]{\strut{}$SM_{QCD+EW}$}}%
\put(12572,5389){\makebox(0,0)[r]{\strut{}$SM_{QCD+EW} + MW$}}%
\end{picture}%
\endgroup
\caption{\label{fig4} NLO decay width, $\Gamma_{2L}(H\to\gamma\gamma)$ in 
the SM and in MW model, normalized to the LO decay width 
$\Gamma_{1L,SM}(H\to\gamma\gamma)$ in the SM.
The dashed line represents the two-loop EW SM corrections; the solid
line represents the two-loop QCD SM corrections. The dotted line represents
the full SM two-loop corrections. Finally, the dashed-dotted line represents
the corrections due to the full two-loop SM+MW decay width.
The MW parameters are chosen as in Fig.~\ref{fig3}}
\ec
\end{figure}

In Fig.~\ref{fig4} we plot  the correction to the decay width $\Gamma(H \to
\gamma \gamma)$ originating from the two-loop (NLO)  corrections, with 
respect to the one-loop (LO) SM prediction, assuming the same 
parameters as in Fig.~\ref{fig3}. In the figure, the EW and
QCD SM corrections are separately shown as well as their sum.
As already pointed out in Refs.~\cite{DM2,Passarino:2007fp}, the SM NLO EW 
and QCD corrections basically cancel each other. The effect of the charged 
scalar particle, on top of the Standard Model particles, is to reduce the 
LO SM decay width into two photons. At the leading order (one-loop amplitudes), 
this reduction ranges already between 9 and 6\%. Considering the QCD 
corrections to the scalar contribution (Fig.~\ref{fig4} dashed-dotted line), 
the reduction is further increased by an amount that ranges between 13\% for 
$\mh \sim 100$~GeV and 6\% for $\mh \sim 160 $~GeV. The reduction effect 
clearly depends on the mass of the scalar particle and can be much more  
pronounced for smaller values of $m_S$. 

It should be recalled that the relevant quantities at LHC for the Higgs 
discovery are the product of the production cross sections times the branching 
ratios, so that for $\mh \lesssim 140$~GeV the relevant quantity is
$\sigma(pp \to H) BR(H \to \gamma \gamma)$. In this product 
the reduction effect induced in $ H \to \gamma \gamma$ by the scalar 
contribution is actually more than compensated by the increase in the 
gluon fusion  production cross section, so that the Higgs boson discovery 
potential at LHC in this model is actually higher than in the SM.

\subsection{The MSSM \label{MSSM}}

We consider now the contribution of the scalar quarks on the Higgs production
cross section in the MSSM. The Higgs sector of the MSSM contains two complex
$({\bf 1},{\bf 2})_{1/2}$, $({\bf 1},{\bf 2})_{-1/2}$ scalar fields that
couple to the down- and up-type fermions separately. After spontaneous
symmetry breaking the spectrum of the MSSM Higgs sector contains five
physical states, two CP-even neutral boson, $h,H$, one CP-odd neutral one, $A$,
and two charged Higgs particles $H^\pm$. At the lowest order the MSSM Higgs
sector can be specified in terms of two independent parameters, usually chosen
as $m_A$, the mass of the pseudoscalar boson, and $\tan\beta = v_2/v_1$,
the ratio of the vacuum expectation values of the two Higgs fields.

Stop and sbottom loops can affect significantly the production cross section 
of the CP-even Higgs bosons. Indeed, there can be regions of the SUSY parameter 
space in which one of these squarks can be relatively
light and its coupling to the $h$ boson relatively strong. As a result
this state does not decouple in the $gg \to h$ amplitude and actually
its contribution interfere with the fermion one. It should be recalled that 
when going to the  NLO level the purely squark contribution is only part of the 
MSSM QCD correction to the production cross section.
Indeed, at this level, besides diagrams containing quarks or squarks and
gluons also diagrams with quark, squark and gluino can contribute and a
clear separation between the two contributions is not possible
\cite{Harlander:2003bb}. However, we are going to consider a scenario in which
one squark is supposed to be relatively light and therefore to provide the
bulk of the corrections while gluino diagrams are supposed to give a small 
contribution that we are going to neglected. Also we are not going to take 
into account the quartic self-interaction coupling among squarks.

The computation of $ gg \to h/H$ cross section in the MSSM requires
the knowledge of the particle mass spectrum of this model. Nowadays, it
is available a set of computer codes \cite{codes1} that allow to
compute the entire MSSM spectrum starting from a restricted number of
parameters, that can be assigned at a high scale and then evolved down
to the weak scale, like for example in a MSUGRA scenario, or directly
assigned at the weak scale. At the level of NLO corrections it is
important to specify exactly the meaning of these parameters. For what
concerns the entries in the squarks mass matrix, the output of these
codes is usually expressed in terms of dimensionally reduced
$\overline{\mbox{DR}}$ parameters evaluated at some specified
$\mu=\mu_{EWSB}$ scale.  Consequently the mass eigenvalues obtained
from this squarks mass matrix, as well as the couplings of the squarks
to the neutral Higgs bosons should be intended as
$\overline{\mbox{DR}}$ quantities\footnote{The codes usually provide
  also OS masses for the SUSY particles.}.  Among all the various
quantities entering in the formulae for the gluon fusion cross section
at NLO only ${\mathcal G}^{(2l,C_R)}_{0}$ requires an exact
specification\footnote{For the top and bottom contribution to the
  gluon fusion cross section we consider always OS masses.}. In
Ref.~\cite{Aglietti:2006tp} this quantity is reported in terms of
dimensionally regularized $\overline{\mbox{MS}}$ masses and
couplings. Thus, to employ the result of Ref.~\cite{Aglietti:2006tp} we
have first to convert the $\overline{\mbox{DR}}$ masses and couplings
obtained as output from any $\overline{\mbox{DR}}$ code into
$\overline{\mbox{MS}}$ quantities. For what concerns the masses one
notices that among the various parameters entering in the squark mass
matrix \be m^2_{\tilde q}=\pmatrix{ { m^2_{\tilde q_L}}+m^2_q + \mz^2
  (I^3_q- e_q \sin^2 \theta_W)\cos 2\beta & m_q( A_q-\mu
  \,(\cot\beta)^{2 I^3_q}) \cr m_q( A_q-\mu \, (\cot\beta)^{2 I^3_q}
  )& m^2_{\tilde q_R}+m^2_q+ \mz^2\, e_q \sin^2\theta_W\cos 2\beta }\,
\label{qmass}
\ee
the soft SUSY breaking  left- and right-handed squark masses, 
$m^2_{\tilde q_L},\, m^2_{\tilde q_R}$, the trilinear squark coupling, $A_q$, 
$\tan \beta$ as well as the Higgs mass parameter $\mu$ at the level of NLO 
QCD corrections are
identical in  dimensional regularization and dimensional reduction while the 
only parameter that requires a conversion is the quark mass as
\be
m_q^{(\overline{\mbox{DR}})} = 
           m_q^{(\overline{\mbox{MS}})} -\frac{g_s^2}{16 \pi^2} C_F \,m_q~.
\label{conversion}
\ee 
In Eq.~(\ref{qmass}) $I^3_q$ is the third component of the weak
isospin, $e_q$ the electric charge of the quark $q$, $\mz$ the mass of
the $Z$ boson and $\theta_W$ the Weinberg angle. Once an
$\overline{\mbox{MS}}$ squark mass matrix has been constructed one can
obtain the $\overline{\mbox{MS}}$ mass eigenvalues and in case
convert them in the OS results. A similar procedure can be employed to
obtain the $\overline{\mbox{MS}}$ couplings of the squarks to the
Higgses \cite{couplings,Muhlleitner:2006wx}, where also in this case the 
only quantity that requires a conversion is the quark mass as in 
Eq.~(\ref{conversion}).

Having set the framework for the computation of the gluon fusion production
cross section of the neutral MSSM Higgs bosons, we consider the  particular
region of the
parameter space, the so-called Higgs gluophobic scenario, in which there is a
negative interference between the standard fermionic contribution and
the one coming from the stop and sbottom states \cite{Djouadi:1998az}.
As input parameters for the squark mass matrix at  the scale 
$\mu_{EWSB}=300$~GeV in this scenario
we chose $m^2_{\tilde q_L} = m^2_{\tilde t_R}= m^2_{\tilde b_R}= 350$~GeV 
$A_t=A_b= -600$~GeV, $\mu = 300$~GeV, 
$\mt^{\overline{\mbox{MS}}}(\mu_{EWSB})=153$~GeV,
$m_b^{\overline{\mbox{MS}}}(\mu_{EWSB})=2.3$~GeV.

\begin{figure}
\begin{center}
\begin{picture}(0,0)%
\includegraphics{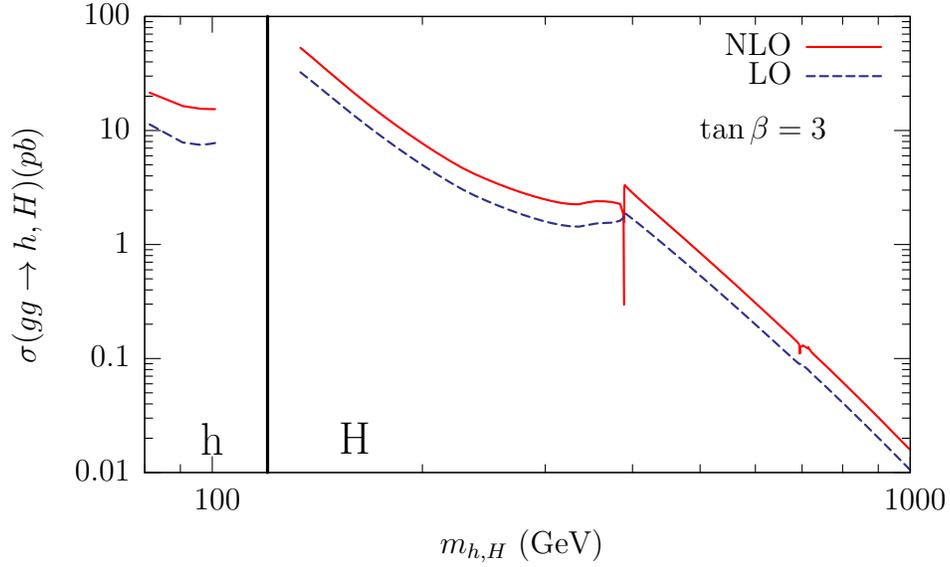}%
\end{picture}%
\begingroup
\setlength{\unitlength}{0.0200bp}%
\begin{picture}(18000,10800)(0,0)%
\put(2475,1650){\makebox(0,0)[r]{\strut{} 0.01}}%
\put(2475,3800){\makebox(0,0)[r]{\strut{} 0.1}}%
\put(2475,5950){\makebox(0,0)[r]{\strut{} 1}}%
\put(2475,8100){\makebox(0,0)[r]{\strut{} 10}}%
\put(2475,10250){\makebox(0,0)[r]{\strut{} 100}}%
\put(4024,1100){\makebox(0,0){\strut{} 100}}%
\put(17175,1100){\makebox(0,0){\strut{} 1000}}%
\put(550,5950){\rotatebox{90}{\makebox(0,0){\strut{}$\sigma(gg\to h,H)(pb)$}}}%
\put(9962,275){\makebox(0,0){\strut{}$m_{h,H}$ (GeV) }}%
\put(4024,2297){\makebox(0,0){\Large h}}%
\put(6709,2297){\makebox(0,0){\Large H}}%
\put(13216,8100){\makebox(0,0)[l]{\strut{}$\tan\beta=3 $ }}%
\put(14950,9675){\makebox(0,0)[r]{\strut{}NLO}}%
\put(14950,9125){\makebox(0,0)[r]{\strut{}LO}}%
\end{picture}%
\endgroup
\caption{\label{fig6a} Production cross section of a
  light (heavy) CP-even Higgs boson, in the MSSM, with $\tan\beta=3$ at
  LO (dashed line) and at NLO (solid line).}
\end{center}
\end{figure}

\begin{figure}
\begin{center}
\begin{picture}(0,0)%
\includegraphics{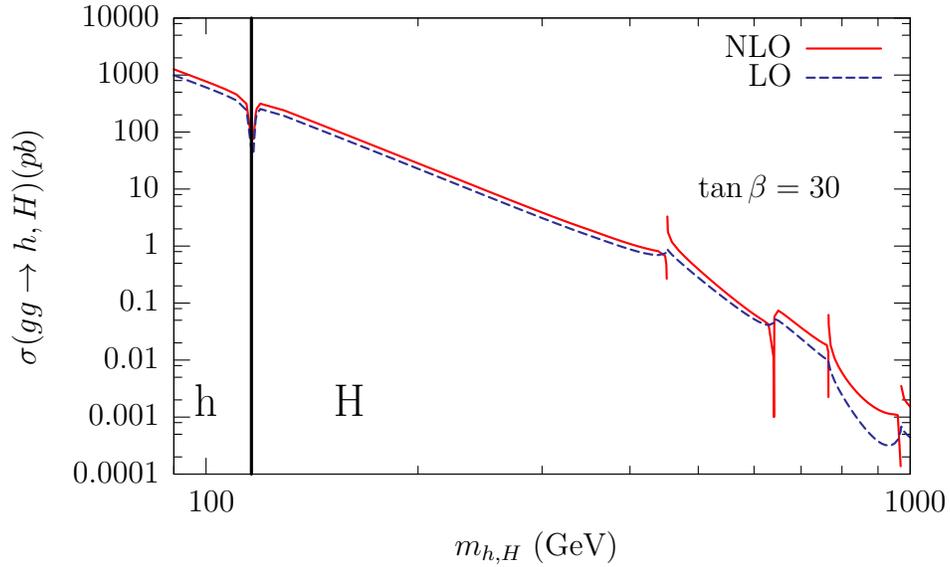}%
\end{picture}%
\begingroup
\setlength{\unitlength}{0.0200bp}%
\begin{picture}(18000,10800)(0,0)%
\put(3025,1650){\makebox(0,0)[r]{\strut{} 0.0001}}%
\put(3025,2725){\makebox(0,0)[r]{\strut{} 0.001}}%
\put(3025,3800){\makebox(0,0)[r]{\strut{} 0.01}}%
\put(3025,4875){\makebox(0,0)[r]{\strut{} 0.1}}%
\put(3025,5950){\makebox(0,0)[r]{\strut{} 1}}%
\put(3025,7025){\makebox(0,0)[r]{\strut{} 10}}%
\put(3025,8100){\makebox(0,0)[r]{\strut{} 100}}%
\put(3025,9175){\makebox(0,0)[r]{\strut{} 1000}}%
\put(3025,10250){\makebox(0,0)[r]{\strut{} 10000}}%
\put(3907,1100){\makebox(0,0){\strut{} 100}}%
\put(17175,1100){\makebox(0,0){\strut{} 1000}}%
\put(550,5950){\rotatebox{90}{\makebox(0,0){\strut{}$\sigma(gg\to h,H)(pb)$}}}%
\put(10237,275){\makebox(0,0){\strut{}$m_{h,H}$ (GeV) }}%
\put(3907,3049){\makebox(0,0){\Large h}}%
\put(6615,3049){\makebox(0,0){\Large H}}%
\put(13181,7025){\makebox(0,0)[l]{\strut{}$\tan\beta=30 $ }}%
\put(14950,9675){\makebox(0,0)[r]{\strut{}NLO}}%
\put(14950,9125){\makebox(0,0)[r]{\strut{}LO}}%
\end{picture}%
\endgroup
\caption{\label{fig6b} Production cross section of a
  light (heavy) CP-even Higgs boson, in the MSSM, with $\tan\beta=30$ at
  LO (dashed line) and at NLO (solid line).}
\end{center}
\end{figure}

In Fig.~\ref{fig6a} we plot the gluon fusion production cross section
for the $h,\,H$ CP-even Higgs bosons at LO and NLO for $\tan\beta =
3$.  The $\overline{\mbox{MS}}$ squark mass eigenvalues are found to
be: $m_{\tilde t_1}= 190$ GeV, $m_{\tilde t_2}= 500$ GeV, $m_{\tilde
  b_1}= 350$ GeV, $m_{\tilde b_2}= 360$ GeV while the rest of the MSSM
particle spectrum, in particular the masses of the lighter and heavier
neutral CP-even Higgs bosons, is obtained using the code Suspect with
a gluino mass $m_{\tilde g}= 500$ GeV and $M_2 = \mu$.  As can be seen
from the figure, when the QCD corrections are taken into account the
NLO cross section shows an increase comparable to the SM
case. Therefore, the QCD corrections to the quark and squark
contributions are both large and of similar size. 
According to the
discussion in Section~\ref{sec3} the NLO curve should contain spikes
in correspondence of the opening of the $2\, \tilde t_{1,2}, \, 2\,
\tilde b_{1,2}$ thresholds. These spikes are actually extremely
narrow and either are not drawn or are just hinted in the figure.  In
Fig.~\ref{fig6b} the same analysis is performed for $\tan\beta = 30$ 
with a corresponding squark mass spectrum $m_{\tilde t_1}= 230$~GeV, 
$m_{\tilde t_2}= 490$~GeV, $m_{\tilde b_1}= 320$~GeV, $m_{\tilde b_2}= 380$~GeV.
As can be seen from the figure, in this case the NLO corrections are usually
percentually smaller than in the SM case. Furthermore, in the
singular behaviour  at the openings of the squarks thresholds it is possible
to appreciate the change of sign when passing through the thresholds due to 
our choices of $\overline{\mbox{MS}}$ masses (see Fig.~\ref{fig1}).
Our results for the MSSM are in agreement with the analysis carried out in
Ref.~\cite{Muhlleitner:2006wx}.
%

\section{Conclusions \label{CONC}}

In this paper we have presented analytical results for the NLO QCD 
corrections to the Higgs production 
cross section in gluon fusion. We considered both the contributions
due to colored fermions and colored scalars running in the loops. 
The analytic formulas are provided in a fully general form and they 
can be used in computer codes aiming at the phenomenological description 
of different theoretical models. In particular, the 
results have been implemented in a {\tt Fortran} code 
which provides a flexible tool to study BSM physics effects
for a generic model which satisfies $SU(N_c)\times SU(2)_L\times U(1)_Y $ 
gauge invariance.

We have discussed the behaviour of the $K$ factor (the ratio between the
NLO and the LO cross section), comparing the exact results with the ones 
in which the NLO corrections are calculated in the infinite fermion and/or 
scalar mass approximation. In particular, we recover the well known fact 
that in the SM the effective theory provides a good approximation of the 
NLO corrections in a wide range of values of the Higgs mass. The scalar 
case, however, is more complicated. Actually, a non-physical singularity 
appears in the two-loop virtual corrections at the scalar pair-production 
threshold (with different shapes depending on the renormalization scheme in
which coupling and masses are renormalized). This alters, to some extent, 
the discussion about the behaviour of the $K$ factor near the threshold, 
where the propagator of the scalar field, in principle, should be resummed to 
all orders in perturbation theory.  In the region away from 
threshold, we see a good convergence to 1 in the light-Higgs region, while 
a certain deviation from the effective theory shows up  for 
heavy Higgs masses, ranging in any cases within 10\%.
We notice that, when there is a substantial interference between the
fermion and  scalar contributions the situation could be
more complicated, giving rise to behaviours that could differ substantially
from the ones described above.

In the paper,
we have applied our results to the study of the Higgs production cross section 
in two different extensions of the SM: the model proposed by Manohar and Wise,
in which the SM is supplemented by an extra colored scalar weak doublet 
in the adjoint representation of $SU(N_c)$, and the MSSM
in the limit of neglecting the gluino contribution.
In the MW model, the extra scalars lead to a large enhancement of the Higgs 
production cross section: with the set of parameters considered, we register 
an enhancement of up to a factor of 2 with respect to the SM results. 
In the same model, the Higgs decay width in two photons is decreased by a 
factor up to 13\%. The net effect, considering the combination of production 
and decay, is a large positive correction.
In the MSSM, we consider the so-called gluophobic scenario in which the
destructive interference between squark and quark loops reduces 
significantly the production of the lightest CP-even Higgs boson $h$.
In this situation we find that the NLO corrections to the squark contribution 
are of  similar size as those of the quark part in 
agreement with previous results in the literature 
\cite{Dawson:1996xz,Muhlleitner:2006wx}.

The study of exclusive observables will be necessary to obtain more realistic 
phenomenological results: the squared matrix elements described in this paper 
can be easily embedded in a Monte Carlo code aiming at such studies.

\subsection*{Acknowledgements}

The authors want to thank P.~Slavich for useful discussions and M.~Spira for 
the careful numerical comparison of the two-loop virtual corrections. R.~B.
wishes to thank G.~Rodrigo and M.~Grazzini for useful discussions, and the
Departments of Physics of the Universities of Roma Tre, Milano and Firenze 
for kind hospitality during a part of this work. The work
of R.~B. was partially supported by Ministerio de Educaci\'on y Ciencia (MEC) 
under grant FPA2004-00996, Generalitat Valenciana under grant GVACOMP2007-156, 
European Commission under the grant MRTN-CT-2006-035482 (FLAVIAnet),
and MEC-INFN agreement. The work of G.~D. and A.~V. was supported by the
European Community's Marie-Curie Research Training Network under contract
MRTN-CT-2006-035505 (HEPTOOLS).


\end{document}